\documentclass{article}
\begin{document}
\title{Einstein-Proca Model, Micro Black Holes, and Naked Singularities}
\author{Chris Vuille\\Department of Physical Sciences\\Embry-Riddle Aeronautical University\\Daytona Beach, Florida 32114 \\vuille@erau.edu \and
James Ipser\\ Department of Physics\\University of Florida\\Gainesville, Florida 33224 \and
Jeff Gallagher\\ Department of Physical Sciences\\ERAU\\Daytona Beach, Florida 32114}

\maketitle
\abstract{ The Einstein-Proca equations, describing a spin-1 massive vector field in general relativity,
are studied in the static, spherically-symmetric case. The Proca field equation is a highly nonlinear wave equation, but can be solved to good accuracy in perturbation theory, which should be very accurate for a wide range of mass scales. The resulting first order metric reduces to the Reissner-Nordstrom solution in the limit as the range parameter $\mu$ goes to zero. The additional terms in the $g_{00}$ metric are
positive, as in Reissner-Nordstrom, in agreement with previous numerical solutions, and hence involve
naked singularities. Note: This paper was published in General Relativity and Gravitation, May 2002.}

\section{Introduction}

An exact solution for the Einstein-Proca system for an idealized
point particle has yet to be found \cite{Tucker1}, \cite{Kramer1}.
Such systems have been occasionally discussed in the literature,
for example in Dereli et al. \cite{Dereli}, and have been invoked
by Tucker and Wang \cite{Tucker2} in connection with dark matter
gravitational interactions, where it was shown that such fields
could explain in part the galactic rotation curves.  Numerical
solutions were found independently by Obukov and Vlachynsky
\cite{Obukov}and Toussaint \cite{Toussaint}.  These latter two
papers demonstrated the existence of naked singularities in this
system.  In this section, the system will be solved up to a final
integral, which will then be subjected to perturbation analysis.

Consider a force modeled as a Proca interaction.  During
gravitational collapse, the equivalent of the force charge,
referred to here as the Proca charge, would not be cancelled by an
accumulation of opposite charges, as in electromagnetic
interactions. The stress energy of the force field would
therefore be expected to make contributions to the gravitational
field of the spacetime surrounding the collapsed object.  Because
both the force and the associated gravitational field fall off
exponentially, the effect on the spacetime surrounding a
stellar-size black hole would be completely negligible.

On the other hand, it is thought that microscopic black holes may
have been created in vast numbers during the Big Bang.  These
micro black holes would be expected to have a variety of different
sizes, including, conceivably, some on the order of a femtometer
across.  For such objects, there is the possibility that
associated fields of  Proca-type would prevent the formation of
event horizons, leaving a (short-lived) naked singularity.  This,
then, might be considered a counter-example to Penrose's cosmic
censorship conjecture.

     The equation for a particle exhibiting a spin-1 short or
intermediate-range field in flat space is  Proca's equation
\cite{Proca}, which in the absence of currents is

\begin{equation}\label{a}
\partial_a F^{ab}+\mu^2 A^b= 0
\end{equation}
where
\begin{equation}\label{b}
F_{ab}=\nabla_a A_b -\nabla_b A_a
\end{equation}
The metric will be taken to have diagonal form $c^2,-1,-1,-1$.
The quantity $\mu$ is  a constant, interpreted as being
proportional to the mass of the field quanta and inversely
proportional to the range of the interaction.

Traditionally, the form of equation \ref{a} was chosen for several
good reasons.  First and foremost, it gives an intuitively correct
answer, which is a potential that rapidly falls  off  as $r$ gets
large.  Second, it can be realized by adding a linear term to
Maxwell's equations.  Third, the equation is covariant, and
finally, a Lagrangian exists, meaning this equation is extremal in
a more general function space.

The Lagrangian density for the classic Proca system is:

\begin{equation}\label{h}
\pounds=\sqrt{-g} \left(\alpha F_{ab}F^{ab}+ \beta A_a A^a\right)
\end{equation}
where  $g$ is the determinant of the metric, and $\alpha$ and
$\beta$ are constants.  Varying this equation with respect to
$A^c$ returns equation \ref{b}, provided that $\beta/2
\alpha=-\mu^2$. It turns out that the last term on the right in
\ref{h} , which distinguishes the standard Proca from Maxwell,
causes considerable difficulties in finding the solution to the
general relativistic problem.  These difficulties are absent in
the Reissner-Nordstrom problem primarily due to the antisymmetry
of $F_{ab}$. Nonetheless, considerable progress can be made, as
will be demonstrated in the next section.

\section{Derivation and Solution of the Field Equations}

The metric for static spherical symmetry can be taken to have the
form
\begin{equation}\label{l}
ds^2=e^{\nu} dt^2-e^{\lambda}dr^2-r^2 \left(d \theta^2+\sin^2
\theta d \phi^2 \right)
\end{equation}
Similar forms can also be written down for plane and hyperbolic
symmetry: all subsequent steps in this paper could equally well be
taken in those two cases . The Proca stress-energy tensor can be
obtained from
\begin{equation}
T_{ab}=-\frac{\alpha_M}{8 \pi} \frac{1}{\sqrt{-g}}
\frac{\delta}{\delta g^{ab}}\sqrt{-g} \pounds
\end{equation}
For a given field, the constant $\alpha_M$ is a parameter that
tells how strongly the stress-energy of the field creates
gravitation.  This gravitational strength is so weak compared to
the other forces that it is impractical to determine
experimentally. Again for convenience, this constant and the
factor of $8 \pi$ shall be rolled into the constants $\alpha$ and
$\beta$.  Applying this formula to equation \ref{h} results in
\begin{equation}\label{j}
T_{ab}=2 \alpha {F_a}^dF_{bd}+ \beta A_a A_b -\frac{1}{2} g_{ab}
\left( \alpha F_{cd}F^{cd}+ \beta A_c A^c \right)
\end{equation}

The Proca stress energy, unlike the Maxwell stress-energy, is not
traceless.  Einstein's equations read
\begin{equation}\label{k}
R_{ab}=\kappa \left(T_{ab}-\frac{1}{2} g_{ab}T\right)
\end{equation}

It is advantageous to recast the Proca equation in terms of
ordinary partial derivatives:
\begin{equation}\label{n}
\frac{1}{\sqrt{-g}} \partial_a \left (\sqrt{-g}
F^{ab}\right)-\frac{\beta}{2 \alpha} A^b=0
\end{equation}
The Proca system corresponds to a choice of
\begin{equation}
\frac{\beta}{2 \alpha}=-\mu^2
\end{equation}
We search for a solution of equations \ref{h}-\ref{n} where
$F_{ab}$ is of the form
\begin{equation}\label{o}
F_{ab}=\left( \begin{array}{cccc} 0&-A_0'
&0&0\\A_0'&0&0&0\\0&0&0&0\\0&0&0&0
\end{array}\right)
\end{equation}

With these choices, the stress-energy tensor becomes
\[T_{ab}=\alpha A_0'^2\left(\begin{array}{cccc} -e^{-\lambda}&0&0&0\\0&e^{-
\nu}&0&0\\0&0&-r^2e^{-\lambda- \nu}&0\\0&0&0&-r^2\sin^2 \theta
e^{-\lambda-\nu} \end{array}\right) +\]
\begin{equation} + \frac{\beta A_0^2}{2} \left(\begin{array}{cccc} 1&0&0&0\\0&e^{\lambda -
\nu}&0&0\\0&0&r^2 e^{-\nu}&0\\0&0&0&r^2 \sin^2 \theta e^{-\nu}
\end{array} \right)
\end{equation}

Einstein's equation then can be written down as
\begin{equation} \label{r}
R_{00}=e^{\nu-\lambda} \left(
\frac{\nu''}{2}-\frac{\nu'\lambda'}{4}+\frac{\nu'^2}{4}
+\frac{\nu'}{r}\right) =-\kappa \alpha {A_0'}^2 e^{-\lambda}+
\kappa \beta {A_0}^2
\end{equation}
\begin{equation}\label{s}
R_{11}= \left(
-\frac{\nu''}{2}+\frac{\nu'\lambda'}{4}-\frac{\nu'^2}{4}
+\frac{\lambda'}{r} \right)=\kappa \alpha {A_0'}^2 e^{-\nu}
\end{equation}
\begin{equation}\label{t}
R_{22}=1+ e^{-\lambda} \left(-1-\frac{r \nu '}{2}+\frac{r \lambda
'}{2} \right)= -\kappa \alpha r^2 {A_0'}^2 e^{-\lambda-\nu}
\end{equation}

Of course, $R_{33}= R_{22} \sin ^2 \theta$. Finally, the equation
\ref{a}for the massive vector field is given by

\begin{equation}
A_0'' + \frac{2}{r} A_0' - \left( \frac{\lambda'}{2} +
\frac{\nu'}{2} \right) A_0' +\frac{\beta}{2 \alpha} e^\lambda
A_0=0
\end{equation}

On the face of it, these equations are not dissimilar to
Einstein-Maxwell, differing only by the inclusion of two rather
innocuous terms.  In fact, these small changes result in a
tremendous complications, as will soon be seen.  In the first
place, unlike Einstein- Maxwell, the enormous simplification of
$\lambda'+\nu'=0$ does not occur.  Indeed, multiplying equation
\ref{r}   by $e^{-\nu+\lambda}$ and adding to equation \ref{s}
yields
\begin{equation}\label{t1}
\frac{\nu'}{r}+\frac{\lambda'}{r}=\kappa \beta {A_0}^2
e^{\lambda-\nu}
\end{equation}

Solving this equation for $\lambda'$ and substituting into
equation \ref{t} results, after some algebra, in:
\begin{equation}
e^{\lambda} = \frac{ 1+ r\nu' -\kappa \alpha r^2 {A_0 '}^2
e^{-\nu}}{1+\frac{1}{2}\kappa \beta r^2 {A_0}^2 e^{-\nu}}
\end{equation}
So the function $e^{\lambda}$ has been solved in terms of the
other two functions.  This result, when substituted into the 00
and 11 equations, makes them identical.  Using the last two
equations, the remaining equations for $\nu$ and $A_0$ can be
written as:
\begin{equation}\label{16}
\nu''+{\nu'}^2+\frac {2 \nu'}{r} = -2 \kappa \alpha {A_0'}^2
e^{-\nu} + \left(2+\frac{r\nu'}{2} \right) \kappa \beta {A_0 }^2
e^{-\nu} \frac{ 1+ r\nu' -\kappa \alpha r^2 {A_0 '}^2 e^{-
\nu}}{1+\frac{1}{2}\kappa \beta r^2 {A_0}^2 e^{-\nu}}
\end{equation}
\begin{equation}\label{17}
A_0'' + \frac{2}{r} A_0' =\frac{\beta}{2 \alpha} A_0
\left(-1+\alpha \kappa r A_0 A_0' e^{-\nu} \right)\frac{ 1+ r\nu'
-\kappa \alpha r^2 {A_0 '}^2 e^{-\nu}}{1+\frac{1}{2}\kappa \beta
r^2 {A_0}^2 e^{-\nu}}
\end{equation}

The equation for $\nu$ can be significantly simplified by the
substitution
\begin{equation}
e^{\nu}=f
\end{equation}
where $f=f(r)$.  Substituting this into equation \ref{16} results
in
\begin{equation} \label{18}
f''+\frac{2}{r} f' = -2 \kappa \alpha {A_0'}^2 +\kappa \beta
{A_0}^2 \left(2+\frac{r f'}{2f}\right) \left[\frac{f+rf'-\kappa
\alpha r^2 {A_0'}^2}{f +\frac{1}{2} \kappa \beta r^2
{A_0}^2}\right]
\end{equation}
Similiarly, in equation \ref{17}:
\begin{equation}
A_0'' + \frac{2}{r} A_0' =\frac{\beta}{2 \alpha} A_0
\left(-1+\frac{\alpha \kappa r A_0 A_0' }{f}\right) \left[
\frac{f+rf'-\kappa \alpha r^2 {A_0'}^2}{f +\frac{1}{2} \kappa
\beta r^2 {A_0}^2}\right]
\end{equation}

It may be there is an exact solution for these two equations,
however finding it would be a matter of experimentation and luck,
given the cubic nonlinearities.  A perturbative approach, on the
other hand, has good chances of success, and can be quite accurate
for reasonable values of the parameters of the theory.  The
procedure involves redefining all quantities so that they are
dimensionless, using naturally-occurring parameters.

First, to get the Proca, it is necessary to define $\alpha$ and
$\beta$.  Let these be
\begin{equation}
\alpha=-\frac{1}{2}\epsilon_0
\end{equation}
and
\begin{equation}
\beta=\mu^2 \epsilon_0
\end{equation}
The quantity $\epsilon_0$ fulfills the same function as the
permittivity of free space in electromagnetism, but in this
context pertains to the Proca interaction. $\mu$ is, of course,
the standard range parameter.  Next, set
\begin{equation}
x= \mu r
\end{equation}
This redefines the r-coordinate in terms of a dimensionless
parameter.  The metric function $f$ is already dimensionless;
however $A_0 $ has dimensions of Joules per Proca charge.  Denote
the Proca charge by $q$, in analogy with electromagnetism.  Next,
set
\begin{equation}
A=su
\end{equation}
where
\begin{equation}
s= \epsilon_0^{-1} q \mu
\end{equation}
The parameter $s$ carries all the units of $A$.  Substitute all
these into the above equations and obtain the following two
equations in terms of dimensionless variables only:
\begin{equation}
\left(u''+\frac{2}{x}u' \right)\left(f+\frac{1}{2}\epsilon x^2 u^2
\right)f=u\left(f+\frac{1}{2}\epsilon u u'
\right)\left(f+xf'+\frac{1}{2}\epsilon x^2 {u'}^2\right)
\end{equation}
\begin{equation}
\left(f''+\frac{2}{x}f' -\epsilon {u'}^2
\right)\left(f+\frac{1}{2}\epsilon x^2 u^2 \right)f= \epsilon u^2
\left(2f+\frac{1}{2}xf'\right) \left(f+xf'+\frac{1}{2}\epsilon x^2
{u'}^2\right)
\end{equation}
where
\begin{equation}
\epsilon=\frac{\kappa q^2 \mu^2}{\epsilon_0}
\end{equation}
is a small, dimensionless perturbation parameter, with
$\kappa=G/c^4$.  For a scale similar to that of the strong force,
the factor $\mu^2$ is quite large, $\approx 10^{30}$, and $\kappa
\approx 10^{-44}$.  The remaining term, $q^2/\epsilon_0$, is
analogous to electromagnetic quantities, where the term would have
magnitude of about $10^{-27}$.  Since the strong force is about
100 times stronger than the electromagnetic force, it follows that
this combination of terms should be around $10^{-25}$ in the case
under consideration.  It appears therefore well justified to
consider $\epsilon$ a small quantity for a wide range of scale.
     The functions $u$ and $f$ may therefore be expanded:
\begin{equation}
f=f_0+\epsilon f_1+\epsilon^2 f_2+..
\end{equation}
\begin{equation}
u=u_0+\epsilon u_1+ \epsilon^2 u_2+...
\end{equation}
Inserting these expressions, the following zeroth order equations
are obtained:
\begin{equation}\label{xx}
\left({f_0}''+\frac{2}{x}{f_0}'\right){f_0}^2=0
\end{equation}
\begin{equation}\label{xx2}
\left({u_0}''+\frac{2}{x} {u_0}' \right){f_0}^2=u_0 f_0
\left(f_0+x{f_0}'\right)
\end{equation}
Equation \ref{xx} has the solution
\begin{equation} \label{y}
f_0= a+\frac{b}{x}
\end{equation}
The second term on the right will be the usual Schwarzschild term,
but will evidently be small, and more appropriately first order.
Hence $b$ will be taken to be zero, with $a=1$, giving Minkowski
space as the lowest order in the metric.  With this choice,
equation \ref{xx2} has the usual flat space solution, which is
\begin{equation}
u_0=c_0 \frac{e^{-x}}{x}+c_1 \frac{e^{x}}{x}
\end{equation}
It is evident that $c_1=0$ in this case.  The first order
equations may be written:
\[\left({u_1}''+\frac{2}{x} {u_1}' \right){f_0}^2+\left(u_0''+\frac{2}{x} u_0'\right)f_0
\left(2f_1+\frac{1}{2} x^2 u_0^2 \right)=\]
\begin{equation}\label{xxx2}
=u_0f_0\left(f_1+x{f_1}'+\frac{1}{2}x^2{u_0'}^2\right)+\left(f_0+x{f_0}'
\right)\left(f_0u_1 +u_0f_1+\frac{1}{2}u_0^2u_0'\right)
\end{equation}
\[\left({f_1}''+\frac{2}{x}{f_1}' -
{u_0'}^2\right){f_0}^2+\left({f_0}''+\frac{2}{x}{f_0}'\right)\left(2f_0f_
1 +\frac{1}{2} x^2 u_0^2\right) = \]
\begin{equation}\label{xxx}
={u_0}^2\left(2 f_0+\frac{x}{2}{f_0}'
\right)\left(f_0+x{f_0}'\right)
\end{equation}
The focus here is on equation \ref{xxx}, which yields the
first-order correction to the metric. The homogeneous solution is
again given by equation \ref{y}, except this time the constant
solution will be discarded and the $b/x$ term retained.  This can
be identified with the standard Schwarzschild term.  In addition,
a particular solution is needed.  After substituting the functions
$f_0$ and $u_0$, the equation for $f_1$ becomes
\begin{equation}
{f_1}''+\frac{2}{x}{f_1}' = c_0^2 \left( 3
\frac{e^{-2x}}{x^2}+2\frac{e^{-
2x}}{x^3}+\frac{e^{-2x}}{x^4}\right)
\end{equation}
The particular solution of this equation is
\begin{equation}
f_{1p}=c_0^2 \left(\frac{1}{2}
\frac{e^{-2x}}{x}+\frac{1}{2}\frac{e^{-2x}}{x^2}+\int
\frac{e^{-2x}}{x} dx\right)
\end{equation}
This expression is positive-definite, which will be important in
the subsequent interpretation. The last term can be integrated by
parts to give a slight simplification, which is
\begin{equation}
f_{1p} = \frac{c_0^2}{2} \left( \frac{e^{-2x}}{x^2} +
\int_x^{\infty} \frac{e^{-2x}}{x^2}dx\right)
\end{equation}

The metric function $e^{\nu}$ , with appropriate renormalization
of the constants, can then be written in the form
\begin{equation}
e^{\nu}= 1-\frac{2MG}{c^2 r} + \frac{q^2 G}{\epsilon_0 c^4} \left(
\frac{e^{-2\mu r}}{ r^2}+ \mu^2 \int_r^{\infty} \frac{e^{-2\mu
r}}{r^2} dr\right)
\end{equation}
In the above equation, it has been assumed that the total
classical energy of the field contributes to the gravitational
field.  In the limit as $\mu \rightarrow 0$, corresponding to an
infinite range for the vector potential,  a Reissner-Nordstrom
spacetime is recovered.

\section{Concluding Remarks}

It is thought that numerous micro black holes
may have been created in the early universe. Those black holes would be expected to
evaporate over time due to emission of thermal radiation.  The
positive Proca terms in the above metric suggest the possibility
that some of these objects might be devoid of event horizons, in
agreement with the earlier numerical solutions of Obukov and
Vlachynsky and Toussaint.

Another interesting property of the above solution is that the
gravitational field is repulsive when the constants take on
suitable values, because as $r$ gets very small the exponential
terms will dominate. One is left to speculate whether such
repulsive effects could prevent complete catastrophic
gravitational collapse.

\section{Acknowledgement}
Vuille remembers, with great appreciation, numerous valuable and entertaining discussions with the late Fred Elston on the subject of this paper.

\end{document}